\documentclass[%
superscriptaddress,
reprint,
nofootinbib,
nobibnotes,
amsmath,amssymb,
aps,
floatfix,
]{revtex4-1}

\usepackage[utf8]{inputenc}
\usepackage{tikz}
\usetikzlibrary{shapes.geometric}
\usepackage{xcolor}
\usepackage{placeins}
\usepackage{verbatim}
\usepackage{graphicx}
\usepackage{dcolumn}
\usepackage{bm}
\usepackage{hyperref}
\usepackage{caption}
\usepackage{subcaption}
\usepackage{float}
\usepackage{soul}

\begin{document}

\title{Alternative  Frenkel liquid  Lagrangian}%
\author{F. A. P. Alves-J\'{u}nior}
\email{francisco.artur@univasf.edu.br} 
\affiliation{Universidade Federal do Vale do S\~{a}o Francisco (UNIVASF), \textit{Campus} Serra da Capivara, Brazil}

\author{A.S. Ribeiro}
\author{ G. B. Souza}
\email{antonius.ribeirus@ifpi.edu.br}
\affiliation{Instituto Federal de Educa\c{c}\~{a}o, Ci\^{e}ncia e Tecnologia do Piauí (IFPI), \textit{Campus} São Raimundo Nonato, Brazil}

\author{José A. Helayël-Neto}
\email{helayel@cbpf.br}
\affiliation{Centro Brasileiro de Pesquisas Físicas (CBPF) }

\date{\today}
\begin{abstract}
 Based on the Caldirola-Canai approach,  we endeavor to investigate a dissipative scalar field theory in  Minkowski space-time. We present its free-particle solutions for complex $\omega^{\mu}$ components. We find three profiles of dispersion relations, two of them support gapped-momentum states. We also present an alternative view of this model, where dissipation acts as a geometric effect, and an effective negative scalar curvature space-time emerges. Finally, we illustrate how the present model could be adapted to describe shear waves in Frenkel liquids. 
 \end{abstract}

\maketitle
\section{Introduction}
\hspace{0.5cm}The dissipation phenomena stands as a recurrent theme in Physics. One of the main features of quantum dissipation is the existence of a non-unique formulation \cite{razavy2005classical,kamenev2011field}; one of them is the Caldirola-Kanai(CK) method, that is used to deal with open quantum systems, to search for coherent states \cite{choi2013analysis,tokieda2017quantum}, to study the classical trajectories for the de Broglie-Bohm with dissipation\cite{mousavi2018quantum}, or even to provide new procedures, as in the case of the time-dependent quantum harmonic oscillator\cite{onah2023quadratic}. 

 In high energy scenarios, dissipation is also present in cosmological models for many different reasons\cite{yang2022viscous}, so as to deal with dark matter effects\cite{bhatt2019viscous}. But, dissipation still yields causality and non-stability problems, as the case of dissipative fluid description by the Landau-Lifshitz and Eckart models\cite{landau1971d,eckart1940thermodynamics}. Other theories such as the Israel and Stewart\cite{israel1976nonstationary,israel1979transient,stewart1977transientp} and still more recent proposals try to deal with this open issue\cite{biswas2023expedition, mitra2022causality}. In general, many relativistic theories that investigate dissipation must obey causality conditions, as carefully discussed in \cite{gavassino2022can}.

In Condensed Matter, the Maxwell-Frenkel\cite{frenkel1947liquids,maxwell1867iv,trachenko2023theory} approach for liquids has gained evidence in distinct areas from Biology to Thermodynamics\cite{liu2023general,bolmatov2022phonon,bolmatov2011liquid,trachenko2015collective}.
In particular, a few years ago, Trachenko and his collaborators proposed a  dissipative $PT$-symmetric  scalar field theory \cite{trachenko2017lagrangian,trachenko2019quantum,baggioli2020field} that opens up the path for exciting ideas and interpretation, as it describes a viscous-elastic fluid propagation as quasi-particles. In this description, $\tau_F$ is a characteristic parameter that measures dissipation. 

As an important consequence of the dissipation process, the quasi-particles are governed by the gap moment dispersion relation (DR) that imposes a minimum value for the wave number,$\frac{1}{2\tau_F}$, denoted by $k_g$.  This 
gap means that particles-waves with $k<k_g$ are liquid-like and do not propagate, and particles 
for $k>k_g$ are solid-like particles and they propagate, as in super-fluid dynamics\cite{yang2017emergence}. In literature, gapped moments states are an interesting effect present many models \cite{ishigaki2021nambu,zhong2022transverse,baggioli2019low,yerin2023dielectric}, and a great variety of systems such as fluids,  strongly coupled plasma\cite{baggioli2020gapped,khrapak2019onset}, black hole physics\cite{tanaka2022dissipation}, and in holographic models \cite{baggioli2019maxwell}.

This manuscript asks: Is the Trachenko model the only way to introduce gapped momenta in scalar field theories? To pursue an investigation on these points, we present a relativistic scalar field Lagrangian using the Caldirola-Kanai elements. We work out a Lorentz invariant dissipation equation, an extended telegraph-type equation. We obtain three profiles of solutions depending on complex eigenvalues $\omega^\mu$ components, for two of them, we find the relativistic gap moment states. We also discuss an equivalence between our model and a free particle traveling in an exotic curved space-time. Finally, we adapt our model to describe the usual shear waves in Frenkel liquids, as the Trachenko model.
   
The present contribution is organized as follows. In Section $II$, we present the Lagrangian we start off from.
In Section $III$, we discuss a geometric interpretation for our model. Section $IV$ discusses three profiles of free particles in this dissipate model. In Section $V$, we discuss the alternative  Frenkel Liquid Lagrangian. In the last Section, we close the paper by casting our Final Considerations.


\section{The Covariant Lagrangian}
The Caldirola-Canai(CK) Lagrangian for a point particle has the  form $L=e^{\beta t}\left(\frac{\dot{x}^2}{2}+V(x)\right)$, where $\beta$ is the frictional parameter. With this in mind, we postulate a $SO(N)$ relativistic scalar field theory expressed in the form  

\begin{equation}\label{eq:lagrangiana}
   \mathcal{L}=e^{\frac{x^{\mu}v_{\mu}}{\tau}}\left[\sum^{N}_{j=1}\frac{\eta^{\mu\nu}}{2}\partial_{\mu}\varphi_{j}\partial_{\nu}\varphi_{j} -V(\varphi_1,...,\varphi_N)\right],
\end{equation}
 for an inertial frame in four-dimensional Minkowski space-time.We use $\eta^{\mu\nu}=diag(+,-,-,-)$. The external field, $v^{\mu}$, represents the observer four-velocity, 
  $\tau$ is the dissipation parameter. Notice that, for $\tau\rightarrow\pm\infty$, there is no dissipation; however, for $\tau\rightarrow 0$, the dissipation increases indefinitely. In the rest frame, the Lagrangian becomes 
  In this way, it is clear that $v^{\mu}$ is the speed of the observer with respect to the rest frame where the Lagrangian reads as below:

  \begin{equation}
\mathcal{L}=e^{\frac{t}{\tau}}\left[\sum^{N}_{j=1}\frac{1}{2}\partial^{\mu}\varphi_{j}\partial_{\mu}\varphi_{j} -V(\varphi_1,...,\varphi_N)\right].
\end{equation} 
 It is possible to see that the mechanical regime is given by  $\mathcal{L}\rightarrow e^{\frac{t}{\tau}}\left(\frac{1}{2}\left[\frac{\partial\varphi }{\partial t}\right]^2 - V \right),$
that is Kaldirola-Canai Lagrangian form. 
The field equations for \eqref{eq:lagrangiana} have a
covariant form described by
\begin{equation}
    \eta^{\mu\nu}\partial_{\mu}\partial_{\nu}\varphi_{j}+\frac{v^{\mu}}{\tau}\partial_{\mu}\varphi_{j}=\frac{\partial V}{\partial \varphi_j}.
\end{equation}
 The second term,$\frac{v^{\mu}}{\tau}\partial_{\mu}\varphi_{j}$, constitutes a four-dissipation of the fields $\varphi_{j}$.  It is important to notice that the dissipation depends on the $v^{\mu}$ direction; for instance, an observer with $v^{\mu}=(\gamma,\gamma\frac{\omega_{r}}{k_r})$ only see free particles, $e^{ik_rx+i\omega_r t}$, traveling to the left, an observer with $v^{\mu}=(\gamma,-\gamma\frac{\omega_r}{k_r})$ detects free particles, $e^{ik_rx-i\omega_r t}$, traveling to the right.

Here, we are particularly interested in the case where  $\varphi=\varphi(x,t)$; as a consequence, the field equations reduce to
\begin{equation}\label{eq:four-dissipation-1}
    -\partial_x^2\varphi_{j}+\partial_t^2\varphi_{j} +\frac{\gamma}{\tau}\partial_t\varphi_{j}
    -\frac{v\gamma}{\tau}\partial_x\varphi_{j}=\frac{\partial V}{\partial \varphi_j},
\end{equation}
we see from this equation that the model is not T-symmetric.
 
\section{A geometric interpretation}

\subsection{Effective geometry}
We now discuss some points of our Frenkel relativistic Lagrangian concerning geometric aspects.
Assuming that $g_{\mu\nu}(v^\alpha)=e^{\frac{v^{\mu}x_{\mu}}{\tau}}\eta_{\mu\nu}$ is an effective metric, we could understand the dissipation as a consequence of the space-time curvature effect. 
This curved space-time geometry depends on the observer's four-velocity. In the literature, metrics that depend on  $v^i$ appear in acoustic space-times\cite{barcelo2004causal} and metrics that depend on the particle energy $p^0$ are present in rainbow gravity models. They are based on the 
construction of a relativity model with two fundamental invariants, $c$, and the Plank length, $l_p$. In our model, $c$ and $\tau$ are the invariants.  Since $\tau$ is a free parameter, if we fix some value for it, then all observers feel the same $\tau$, this is what we mean by $\tau-$invariance.

To this effective geometry, we find a non-vanish effective Riemann tensor. Here, we show the effective Ricci tensor components 
\begin{equation}
  R^{eff}_{\mu\nu}= \begin{pmatrix}
\frac{\gamma^{2} v^{2}}{2 \, \tau^{2}} & -\frac{\gamma^{2} v}{2 \, \tau^{2}} & 0 & 0\\
 -\frac{\gamma^{2} v}{2 \, \tau^{2}} &  \frac{\gamma^{2}}{2 \, \tau^{2}} & 0 & 0\\
0 & 0&  \frac{1}{2 \, \tau^{2}} & 0 \\
0 & 0& 0& \frac{1}{2 \, \tau^{2}} 
\end{pmatrix},
\end{equation}
 and the Ricci scalar,
\begin{equation}
   R^{eff}= -\frac{3\gamma^2}{2\tau^2}e^{\frac{\gamma}{\tau}\left(t-v x\right)}.
\end{equation}

This scalar curvature is everywhere negative, as an anti-de Sitter space-time, and it has an unusual form. In addition to that, the $\tau$ parameter is responsible, at the same time, for curving the space as a consequence of the dissipation process. It is interesting to note that $R^{eff}_{\mu\nu}$ and $R^{eff}$ also depend on the Lorentz factor.  In other words, for a class of frames with the same $v$ and $\tau$, the space-time curvature is the same. However, with the scalar curvature term, we cannot see that for frames with $v=\pm\frac{\omega_r}{k_r}$, as we already stated there is no dissipation.

\subsection{Conformal transformation}
In scalar-tensor theories, the conformal transformation
are used to transform the Jordan (Brans-Dicke like the-
ories) and Einstein (General Relativity) frames. Here
we can make a parallel with these ideas. Given a scalar
field theory in Minkowski space-time described by the
Lagrangian
\begin{equation}
    \label{eq:lagrangeano 2b}
    \mathcal{L}=\frac{1}{2}\partial^{\mu}\varphi\partial_{\mu}\varphi -U(\varphi),
\end{equation}
where $U(\varphi)$ is the potential. Performing the conformal transformation $\eta\rightarrow e^{\psi}\bar{\eta}$,

\begin{equation}\label{eq:conforme}
    \mathcal{L}=\frac{e^{2\psi}}{2}\partial^{\mu}\varphi\partial_{\mu}\varphi -U(\varphi).
\end{equation}
 
 In other words, our theory can be interpreted as a conformal flat scalar field theory, where the conformal field $\psi$ is $\frac{v^{\mu}x_{\mu}}{\tau}$. Therefore, in this point of view, two observers with different $\tau$ and $v$ values are in a different conformal geometry.

\section{The spinless particles}
 In this section, we show that the scalar particles from our model, that live in this geometry are not stable. Once gapped moment particles found in \cite{trachenko2017lagrangian} have the real and a imaginary $\omega$ component, here, we  use the particular eigenvalues $\omega=\omega_r+i\omega_i$ and $k=k_r+ik_i$, or  just $\omega^{\mu}=\omega_{r}^\mu+i\omega_{i}^{\mu}$. The solutions are listed below.

\subsection{Decaying states}
Here we try a relativistic version of a particular solution found in Trachenko papers. We take $\omega=\omega_r+i\omega_i$ and $k=k_r$, and we find states for $\omega_r\neq 0$, that follow the dispersion relation
\begin{equation}\label{eq:novo-omega}
    \omega^{\pm}_r=\pm\frac{1}{\sqrt{2}}\sqrt{{(k_r^2-\frac{\gamma^2}{4\tau^2})}+ \sqrt{{(k_r^2-\frac{\gamma^2}{4\tau^2})^2}+\frac{k_r^2v^2\gamma^2}{\tau^2}}},
\end{equation}
or rewriting using  the transformation $\omega\rightarrow\gamma\frac{\bar{\omega}}{\tau}$, $k\rightarrow\gamma\frac{\bar{k}}{\tau}$,
\begin{equation}\label{Dispersion-caso1-redu}
     \bar{\omega}^{\pm}_r=\pm\frac{1}{\sqrt{2}}\sqrt{{(\bar{k}_r^2-\frac{1}{4})}+ \sqrt{{(\bar{k}_r^2-\frac{1}{4})^2}+\bar{k}_r^2v^2}},
\end{equation}
that is very similar to Trachenko's second model dispersion relation in \cite{trachenko2017lagrangian}, given by the Lagrangian $\mathcal{L}=\frac{\partial\varphi_1}{\partial t}\frac{\partial \varphi_2}{\partial t}
    - \frac{\partial \varphi_1}{\partial x}\frac{\partial \varphi_2}{\partial x}
    +\frac{1}{2\tau}\left(\varphi_1\frac{\partial\varphi_2}{\partial t}
    -\varphi_2\frac{\partial \varphi_1}{\partial t}\right)
  -  \frac{l}{2\tau^2}\left( \varphi_1\frac{\partial\varphi_2}{\partial x}-\varphi_2\frac{\partial\varphi_1}{\partial x}\right)$.

\begin{figure}[h]
    \includegraphics[scale=0.5]{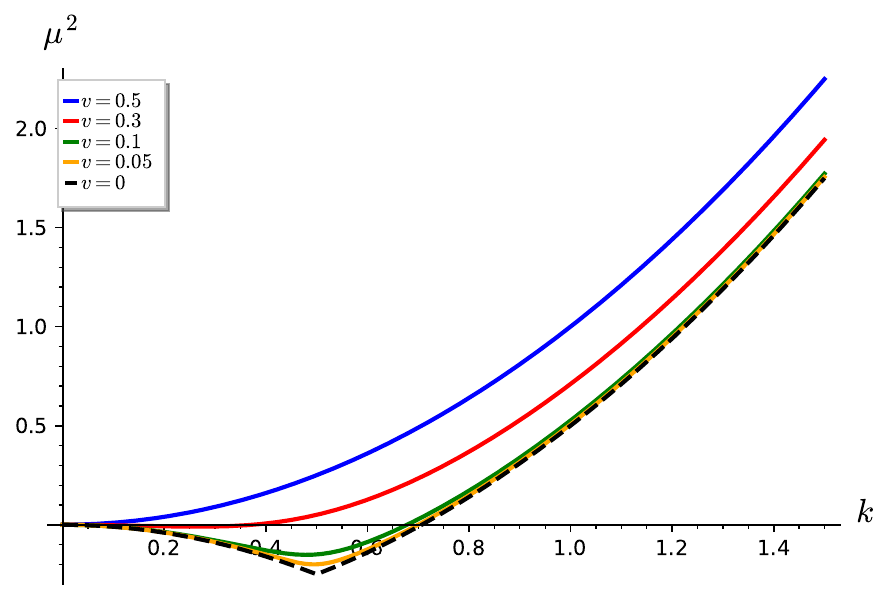}
    \caption{\small The effective mass, $\mu^2$, against $k$ for \eqref{eq:massa-quadrada}.Some observers can see the tachyon-bradyon transition mass.}
    \label{fig:caso-massa}
\end{figure}

\begin{figure}
\includegraphics[scale=0.5]{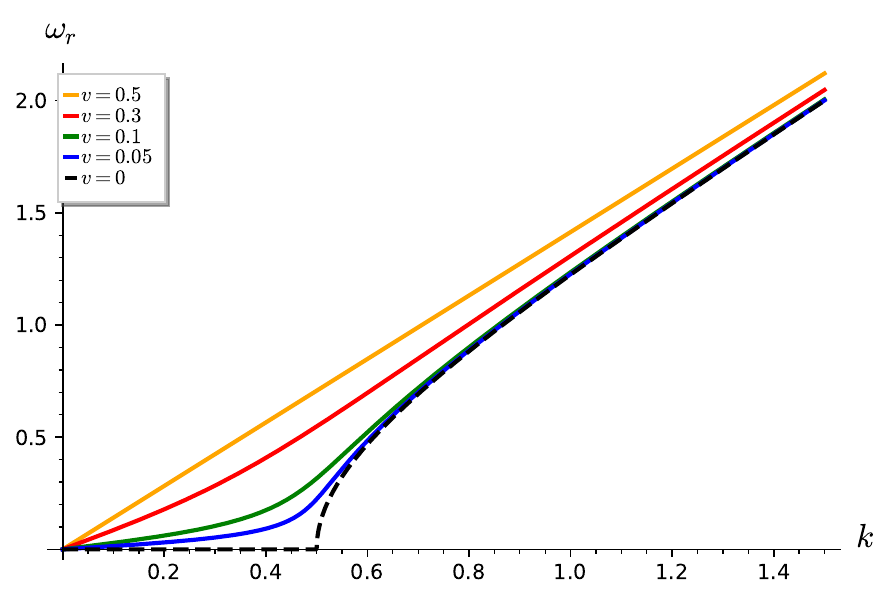}
\caption{\small Dispersion relations for \eqref{Dispersion-caso1-redu}. There is no $k-$gap for $v>0$. This is a relativistic correction to the case $v=0$.}\label{case1}
\end{figure}

From the graph \eqref{case1},  we see distinct profiles for DRs depending on the observation speed $v$. We notice that, for $v\rightarrow 0$(dashed graph), there are k momentum states and some pathological states when $\omega_r= 0$, since for these particles $\omega_i$ is not well defined.  In  order to understand  the relation \eqref{Dispersion-caso1-redu}, we define an effective mass as 

\begin{equation}\label{eq:massa-quadrada}
    \mu^2=-\frac{1}{4}+ \sqrt{{(\bar{k}_r^2-\frac{1}{4})^2}+{\bar{k}_r^2v^2}}.
\end{equation}
We can see that some particles have negative square effective mass(fig. \eqref{fig:caso-massa}); these states occur for frames with $v<\frac{\sqrt{2}}{2}$, and the k-interval $(0,k=\frac{1}{\sqrt{2}}\sqrt{1-2v^2})$, particles with other k-states have $\mu^2\geq 0$. For $k=0$ or $k=\frac{1}{\sqrt{2}}\sqrt{1-2v^2}$, the particles are massless.  For frames with $v>\frac{1}{2\sqrt{2}}$, there is no states with $\mu^2<0$.

If we introduce mass in the field theory, using $V=\frac{m^2\phi^2}{2}$ and $m\rightarrow \frac{\gamma}{\tau}\bar{m}$, then  $\bar{\omega}^{\pm}_r=\pm\frac{1}{\sqrt{2}}\sqrt{{(\bar{k}_r^2-m^2-\frac{1}{4})}+ \sqrt{{(\bar{k}_r^2-\bar{m}^2-\frac{1}{4})^2}+4\bar{k}_r^2v^2}}$; however, there is no qualitative new effect.  For  particles, with $\mu^2 > 0$ we could  associate a Compton wavelength
\begin{equation}
\lambda_C=\sqrt{\frac{1}{{\sqrt{{(\bar{k}_r^2-\bar{m}^2-\frac{1}{4})^2}+4\bar{k}_r^2v^2}-(\bar{m}^2+\frac{1}{4})} }}.
\end{equation}


\begin{figure}\label{fig:vg-solucao-b}
    \centering
    \includegraphics[scale=0.5]{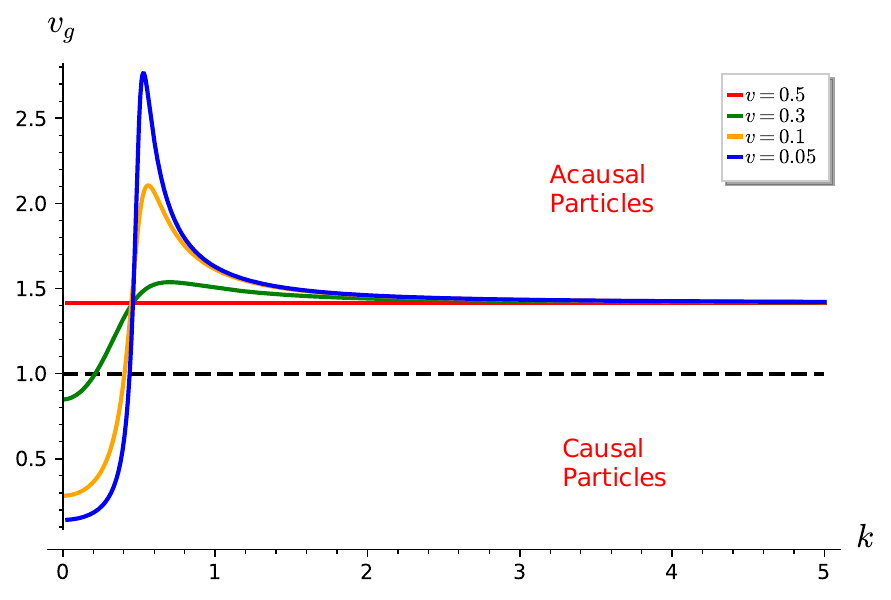}
    \caption{\small For some $k$, we have acausality, see above the dashed line. However, $v_g$ gets limitated as $k\rightarrow\infty$.}
\end{figure}

 Assuming $k_rv>0$,  the dissipated energy is given by 
$\omega_i=\frac{\gamma}{2\tau}\left(1+\frac{vk_r}{\omega^{\pm}_r}\right)$, and the particle/anti-particle solutions are ,respectively,
\begin{equation}
     \varphi_{j}(x,t)=A_0e^{i\omega^{+}_r t+ik_rx}\exp \left(-\frac{\gamma }{2\tau}\left[1+ \frac{ k_rv }{\omega^{+}_r}\right]t\right).
\end{equation}
\begin{equation}
     \varphi_{j}(x,t)=A_0e^{i\omega^{-}_r t+ik_rx}\exp \left(-\frac{\gamma }{2\tau}\left[1- \frac{ k_rv }{|\omega^{-}_r|}\right]t\right).
\end{equation}
Particularly,  the anti-particle wave solution is valid, since  $v<\frac{|\omega_{r}^{-}|}{k_r}$.  Independently of the case, the solutions satisfy the condition  $\varphi(x,t\rightarrow \pm \infty)=0$, which means that the particle must decay. The mean lifetime of the particles depends on the frame's speed. Thus for particles, the mean lifetime is
\begin{equation}
    t_{d}=\frac{2\tau}{\gamma}\frac{\frac{\omega^{+}_r}{k_r}}{\left(\frac{\omega^{+}_r}{k_r}+v\right)},
\end{equation}
 however for the anti-particles, $v<\frac{|\omega^{-}_r|}{k_r}$, this time is
\begin{equation}
    t_{d}=\frac{2\tau}{\gamma}\frac{\frac{|\omega^{-}_r|}{k_r}}{\left(\frac{|\omega^{-}_r|}{k_r}-v\right)},
\end{equation}

When $v=-\frac{\omega^{+}_r}{k_r}$, the particles do not dissipate,  we have a plane wave solution, and $\omega_i=0$, and $\omega_r=k_r$. The same effect happens for anti-particles when $v=\frac{|\omega^{-}_r|}{k_r}$.

\begin{figure}
    \centering
    \includegraphics[scale=0.5]{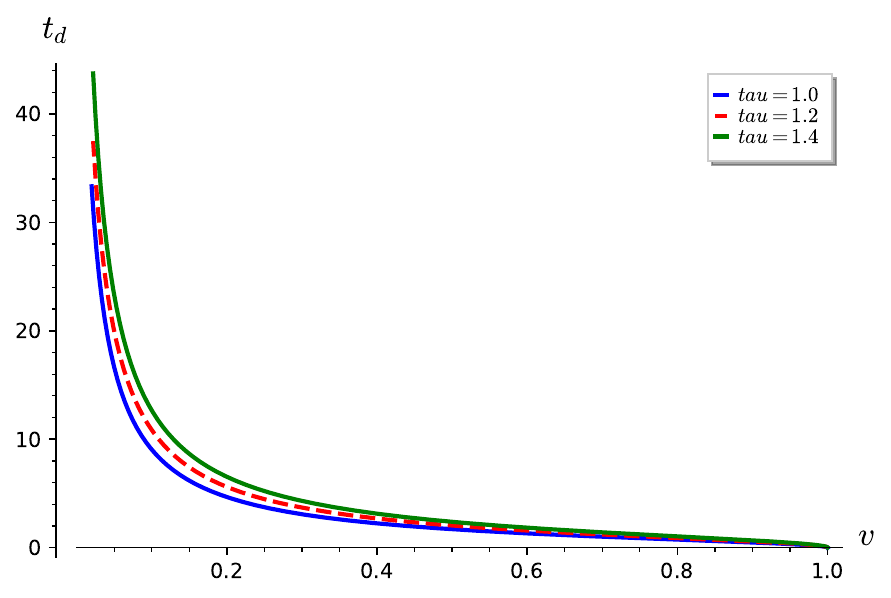}
    \caption{\small We see the mean lifetime for the decay sates $(t_d)$ against the boost
    speed $v$, for different $\tau$ parameters.}
    \label{fig:my_label}
\end{figure}

\subsection{Analogue skin depth effect}
    
Here we present high-energy particles based on \cite{trachenko2021dissipation} solutions. Now assuming that $\omega=\omega_r$ and $k=k_r+ik_i$, then  the scalar field is $\phi=A_0e^{k_ix}\exp\left(i\omega_r^{\pm}t+ik_rx\right)$, where $k_i$ is 
   
    \begin{equation}
        k_i=-\left(v-\frac{\omega^{\pm}_r}{k_r}\right)\frac{\gamma}{2\tau},
        \end{equation}
From the last equation, we see that, in this case, we have a frame-dependent attenuation, and there is a special frame with $v=\frac{\omega_r^{\pm}}{k_r}$, we have free particles. The dispersion relation, $\omega_{r} = |k_{r}|\sqrt{\frac{v^2\gamma^2+4k_r^2\tau^2}{\gamma^2+4\tau^2k_{r}^2}}$. Performing $\omega\rightarrow\gamma\frac{\bar{\omega}}{\tau}$ and $k_r\rightarrow\gamma\frac{\bar{k}_r}{\tau}$, we simplify the DRs, in the form
    \begin{equation}\label{bar-equation-k}
      \bar{\omega}_{r} = |\bar{k}_{r}|\sqrt{\frac{v^2+4\bar{k}_r^2}{1+4\bar{k}_{r}^2}},
    \end{equation}
that shows no k-gap or energy gap. The particle/anti-particle solutions of the equation  are 
 
    \begin{equation}
     \varphi_{j}(x,t)=A_0e^{i\omega_r^{+} t-ik_rx}\exp \left(-k_ix\right)
\end{equation}
\begin{equation}
     \varphi_{j}(x,t)=B_0e^{i\omega_r^{-} t-ik_rx}\exp\left(-k_ix\right).
\end{equation}
for $t>0$ 
These solutions represent attenuated wave solutions or absorbed particles and satisfy the boundary condition $\varphi(x\rightarrow\pm \infty,t)=0$. We see that the penetration depth 
\begin{equation}
\delta_{\pm}=\frac{2\tau}{\gamma \left[ v -\frac{\omega_r^{\pm}}{k_r}\right]}.
\end{equation}
 If $v\geq\frac{\omega^{\pm}_r}{k_r}$, then there is no attenuation phenomena for this frame and we have a plane wave solution.

   \begin{figure}
       \centering
       \includegraphics[scale=0.5]{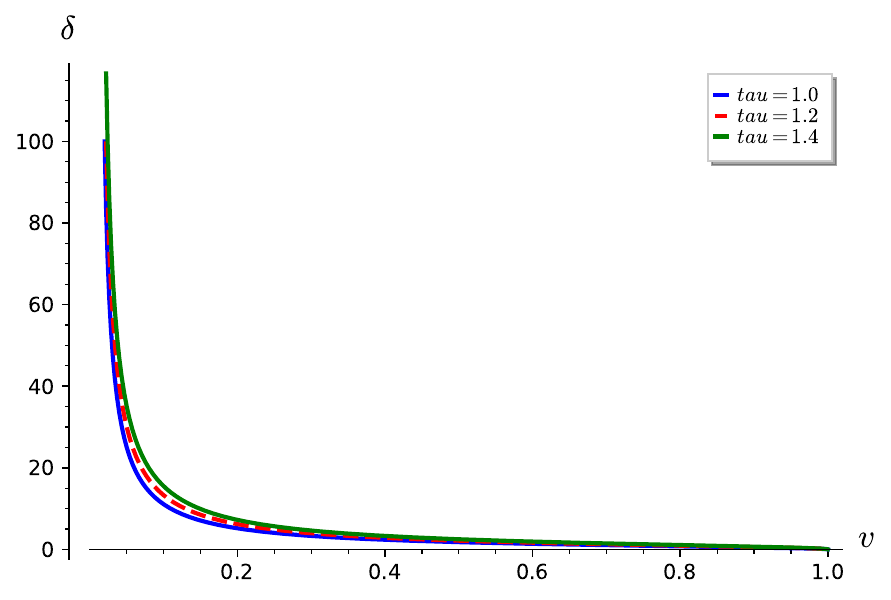}
       \caption{\small{ Here we show the penetration depth $(\delta)$ against the boost speed $v$. We see three different curves for each $\tau$ parameter.}}
       \label{fig:my_label}
   \end{figure}{}

\begin{figure}
    \centering
    \includegraphics[scale=0.5]{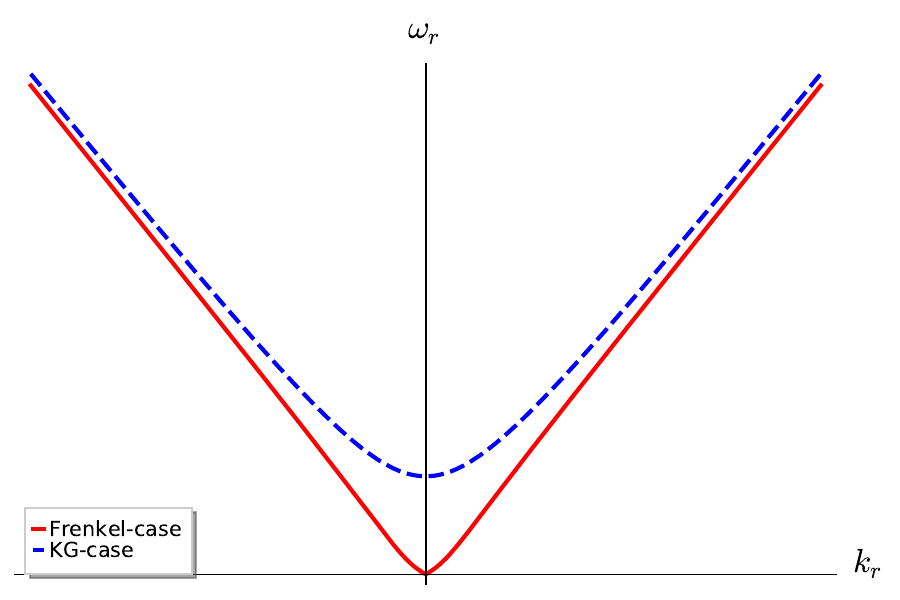}
    \caption{\small The dashed curve(blue) is the dispersion relation for a Klein-Gordon massive particle. The other curve(red) shows the dispersion relation for \eqref{bar-equation-k}.  }
    \label{fig:my_label}
\end{figure}
\begin{figure}
    \centering
    \includegraphics[scale=0.5]{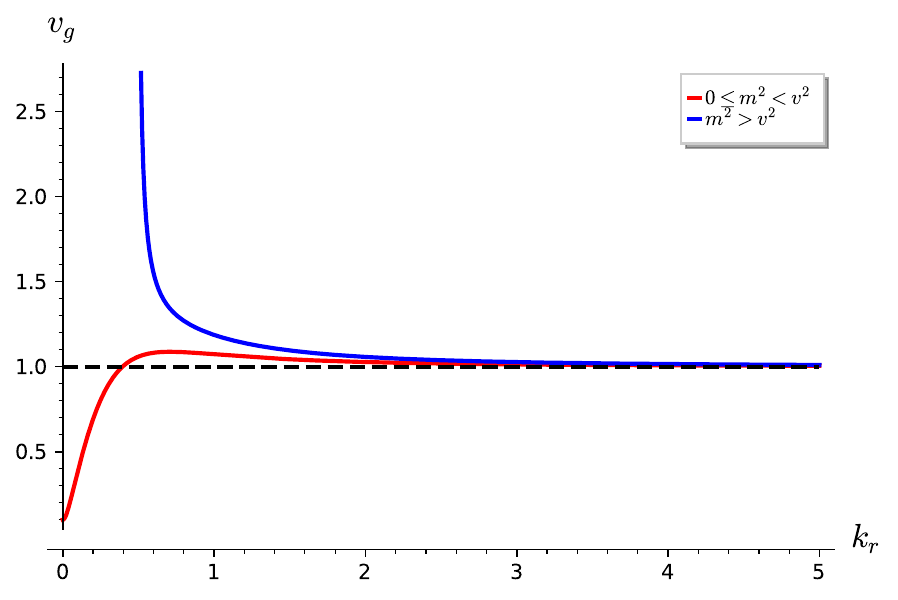}
    \caption{\small The group velocity for particles with and ${\bar{m}^2>v^2}$(blue) and ${\bar{m}^2<v^2}$(red). In this last case, for some $k$ values $v_g$ does not exceed $c$.}
    \label{fig:my_label}
\end{figure}

An interesting dispersion relation \eqref{bar-equation-k} comes out for a massive particle,
\begin{equation}
\bar{\omega}_{r} = |\bar{k}_{r}|\sqrt{\frac{v^2+4\bar{k}_r^2-4\bar{m}^2}{1+4\bar{k}_{r}^2}},
\end{equation}
 where $\bar{m}=\frac{\tau m}{\gamma}$. As we can notice massive particles with $\bar{m}^2>\frac{v^2}{4}$ then we have a k-gap, $k_g=\sqrt{4\bar{m}^2-v^2}$, which means that different frames have distinct $k_g$. Furthermore, by adding a mass to the field the associate particles still have the limit, $\omega_r(k_r\rightarrow0)=0$. 
\subsection{Double effects: dissipation and absorption}
Taking $v>0$, the particles, $\phi=e^{ikx-i\omega^{+} t}$, have their eigenvalues as $\omega^{+}=\omega^{+}_r-i\frac{\gamma}{2\tau}$ and $k=k_r+i\frac{v\gamma}{2\tau}$, and anti-particles,  $\phi=e^{ikx+i\omega^{-} t}$, have different eigenvalues, $\omega^{-}=\omega^{-}_r+i\frac{\gamma}{2\tau}$ and $k=k_r+i\frac{v\gamma}{2\tau}$.  Both types  particles follow  the dispersion relation  $-\omega_r^2+k_r^2-\frac{1}{4\tau^2}=0$, as a consequence, the gap moment equation is
\begin{equation}
\omega^{\pm}_r=   \pm \sqrt{k^2_r-\frac{1}{4\tau^2}},
\end{equation}
where  $-\frac{1}{4\tau^2}$ is the effective square mass term, $\mu^2$. This is the relation is exactly the same of \cite{trachenko2017lagrangian}, however here we are in a relativistic scenario.   
It is essential to notice that, the group velocity,$v_g$, for these particles, exceeds the light speed in vacuum, as a tachyon. The particle/anti-particle  solutions are respectively 
\begin{equation}\label{eq:sol-a1}
    \varphi_{j}(x,t)=A_0e^{-i\omega^{+}_r t+ik_rx}\exp\left(-\frac{\gamma }{2\tau}\left[t+ vx\right]\right),
\end{equation}
\begin{equation}\label{eq:sol-a2}
    \varphi_{j}(x,t)=A_0e^{-i\omega^{-}_r t+ik_rx}\exp\left(-\frac{\gamma }{2\tau}\left[t+ vx\right]\right).
\end{equation}

These dissipative-attenuated solutions are valid for $t+vx>0$. Otherwise, for $t+vx<0$ they grow indefinitely. The effects of dissipation break the T symmetry, as we see from the equation \eqref{eq:four-dissipation-1}. If we add a mass term, $V(\phi)=\frac{m^2}{2}\varphi^2$, the dispersion relation  becomes $\omega^{\pm}_r=   \pm \sqrt{k^2_r-\frac{1}{4\tau^2}+m^2}$.  If $m>\frac{1}{2\tau}$, there is no gap,  and $v_g<c$ and we can associate a  Compton wave length, $\lambda_C=\frac{1}{\sqrt{m^2-4\tau^2}}$. If $m=\frac{1}{2\tau}$, then $v_g=c$ and $\lambda_C\rightarrow\infty$. Thus, we only have a gap for particles where $m^2>\frac{1}{4\tau^2}$ and $v_g>c$ but no $\lambda_C$ could be defined. 

\section{Application: alternative field theory for Frenkel liquid dynamics}

 By Frenkel dynamics, we mean the fact that fluids are generally treated using the hydrodynamics approach,  in this way, the collective
transverse oscillation could not be explained. However, Frenkel's \cite{frenkel1955kinetic} ideas 
led to the description of the  "solid-like" proprieties for liquids.  As a consequence it is possible to calculate liquid heat capacity \cite{bolmatov2012phonon}  in the neighborhood of crystallization point are approximately equal to their corresponding solid heat capacity and understand the superfluids proprieties \cite{yang2017emergence}. 

 There is already a field theory to describe quasi-particles in Frenkel fluids. However, here we adapt our theory in a non-relativistic version as an alternative to the Trachenko model. Our Lagrangian  is
\begin{equation}\label{eq:lagrangeano 2}
    \mathcal{L}=e^{\frac{t}{\tau}}\left[\sum_{j=1}^{N}\frac{1}{2}\eta^{\mu\nu}_{NR}\partial_{\mu}\varphi_j\partial_{\nu}\varphi_j -V(\varphi_1,..,\varphi_N)\right],
\end{equation}
where 
\begin{equation}
    \eta^{\mu\nu}_{NR}=\begin{pmatrix}
1 & 0\\
 0 & -1 
\end{pmatrix}.
\end{equation}
It is clear that the above Lagrangian is not time-reversible as \eqref{eq:lagrangiana} and has its form in the rest frame. The field equations are
\begin{equation}\label{eq:klein1}
    -\partial_{x}^2\varphi_{i}+\frac{1}{\tau}\partial_{t}\varphi_{i}+\partial_{t}^2\varphi_{i}=0,
\end{equation}
that are the same of \cite{trachenko2017lagrangian}. We intend to present a more general profile of eigenfunctions, $e^{i\omega t +ikx}$, and to achieve this we use   $\omega=\omega_r+i\omega_i$, and $k=k_r+ik_i$, and deduce at the dispersion relation
\begin{equation}
    \omega^{\pm}_r=\pm\sqrt{k^2-\frac{1}{4\tau^2}}, \hspace{1cm} \omega_i=-\frac{i}{2\tau},
\end{equation}
that is found in \cite{yang2017emergence}. If we add a mass term to the field, the frequency gets the form $\omega_r=\sqrt{k^2-\frac{1}{4\tau^2}+m^2}$. The  particle solutions are $\varphi_{j}(x,t)=A_je^{\frac{-t}{2\tau}}e^{i\omega_r^{+}t-ikx}$, 
 and the antiparticle solutions are $\varphi_{j}(x,t)=B_je^{\frac{-t}{2\tau}}e^{i\omega_r^{-}t-ikx}$.

 Trachenko's Lagrangian is modeled with  $\phi_1$ and $\phi_2$, which are related to the interactions between the system $1$ and  $2$. Both models describe the waves in Frenkel liquids, however since our result reproduces the $\phi_1=e^{\frac{-t}{2\tau}}\cos(kx-\omega t)$  shear-wave solution present in \cite{trachenko2017lagrangian,trachenko2019quantum,trachenko2021dissipation}, and does not reproduce the PT-reverse shear-wave solution, $\phi_2= e^{\frac{t}{2\tau}}\cos(kx-\omega t)$,
then there is no equivalence between our theories. 

\section{Final Considerations}

 In this contribution, we have presented our efforts to build up a relativistic Frenkel-liquid dynamical model and an alternative formulation to Trachenko's description of shear-waves in liquids. Both models violate the T-symmetry.  We found three profiles of high-energy particles depending on $\omega^\mu$. The dissipative solutions exist whenever $\omega$ has imaginary components and they have no k-gaps. Attenuated solutions show up whenever $k$ has imaginary components and there is k-gap in this situation, if we take $V=\frac{m^2\phi^2}{2}$ as a potential. Finally, we find the dissipative-attenuated particle/antiparticle solution, that also exhibits a k-gap. 
 
We present two views of our model. The first one is the dissipation process for scalars in Minkowski space-time, and a free scalar particle in an exotic curved space-time, where the curvature depends on  $\tau$, the observer's speed, $v$, and the corresponding Lorentz factor.

Following our efforts to handle these scenarios, we could also propose modified models, as in the case
to deal with non-extensivity we can use  $q-$exponetial instead of CK exponentials, likewise it is studied in \cite{choi2014effects}. We also could propose alternative coupling,$e^{\frac{x^{\mu}v_{\mu}}{\tau}}$, terms to investigate the violation of P-symmetric, for example, as a particular scalar field model, $\mathcal{L}=e^{\frac{\gamma}{\tau}\left(x-vt\right)}\left(\partial_{\mu}\varphi\partial^{\mu}\varphi -V\right)$, that is non-P-symmetric  $\partial^{\mu}\partial_{\mu}\varphi +\frac{\gamma}{\tau}\partial_{x}\varphi-\frac{\gamma v}{\tau}\partial_{t}\varphi=0$.

With the same spirit, our results seem to describe a qualitative behavior, similar to the fig.1 of the article \cite{trachenko2017lagrangian}, where the dispersion curves are shown. We think that further investigation should clarify if our curves could really describe the gap-momentum states and the near-gap-momentum states worked out by Trachenko figure.     

The description of the dissipation process in relativistic fluids, quantum mechanics or field-theoretic models is still an open question. There is intensive research on this topic including questions like causality and dissipation. Our solutions deal with   Gavassino's work \cite{gavassino2022can}, particularly with his first theorem, which, in essence, states that, if a perturbation is dissipating in one frame, it is dissipating in another frame, since its propagation is subluminal. At first sight, we see that in the cases $(a)$ and $(c)$, the dissipation rate, $\Gamma$, defined as $(\Gamma=\frac{d \ln \varphi}{d t})$ is $\frac{-\gamma}{2\tau}$ for a moving observer and $\frac{-1}{2\tau}$ for in the rest frame. From a relativistic perspective, all the solutions are in a four-dissipative process, and further analysis is required to check if his theorem is in agreement with our case, once our equations are an extension of the telegraph equation. The connection between the fact that this dissipation rate changes from rest to moving observers, and the analysis of Lorentz invariance for unstable particles in \cite{gavassino2022boosting} are issues that should be clarified. 

On the basis of what we have investigated and reported in this paper, we have in mind new paths of investigation. We have in project to proceed further by studying dissipation in two scenarios. The first one would concern fermionic (spin one-half) fields, by comparing the cases of Weyl (chiral fermions), Dirac and Majorana fields. This could be of interest in connection with aspects of Neutrino Physics \cite{guzzo2016quantum,de2023neutrino} such as, for example, the mechanism of neutrino family oscillations and an inspection of the Pontecorvo-Maki-Nagakawa-Sakata (PMNS) neutrino mixing matrix in connection with dissipation . The second line of investigation we intend to pursue includes the aspects and peculiarities of dissipative field-theoretic models in different space-time dimensions; actually, in low-dimensional systems, namely, (1+1) and (2+1) dimensions, to discuss gapped-momentum states in different (lower) dimensions and to contemplate other planar condensed-matter systems\cite{dattagupta2021spin}. 

\begin{table}[H]
\begin{center}
\renewcommand{\arraystretch}{1.2}
\begin{tabular}{|c|c|c|c|c|}

\hline
\small $\omega^\mu$ & k-gap  &$v_g>c$  \\
\hline                    
\small Complex $\omega$ & No & For some $k$ values\\
 \hline
\small Complex $k$  &For $\bar{m}^2>v^2$ & For some $k$ values  \\
 \hline
\small Complex $k$ and $\omega$ & For $k_r>\frac{1}{2\tau}$ & Always\\
 \hline
 
\end{tabular}
\end{center}
\label{tab1}
\caption{Summary of our wave-particles proprieties.}
\end{table}

Finally, we state that further analysis using the second quantization of our field model should be considered to investigate the quantum scenario and compare it with \cite{trachenko2019quantum}. Also in comparison with Trachenko models, it becomes evident that $\tau \rightarrow \tau_F$. Once $\tau_F$ has temperature dependence, and $\tau$ is an invariant, then the temperature must be an invariant in our models. This is in agreement with Landsberg-Matsas \cite{landsberg2004impossib} point of view in the debate of relativistic temperature \cite{farias2017temperature}. 
\section{Acknowledgements}
I would like to thank R.J.A. Macêdo for useful discussions and for  G.V. Silva critical reading of the paper.
\bibliography{refs}

\end{document}